# Enhancing UX Research Activities Using GenAI – Potential Applications and Challenges

Stefan Graser, Anastasia Snimshchikova, Stephan Böhm

{stefan.graser, stephan.boehm}@hs-rm.de
CAEBUS Center of Advanced E-Business Studies
RheinMain University of Applied Sciences
Wiesbaden, Germany

Martin Schrepp

martin.schrepp@sap.com
SAP SE
Walldorf, Germany

*Abstract*—User Experience (UX) Research covers various methods for gathering the users' subjective impressions of a product. For this, practitioners face different activities and tasks related to the research process. This includes processing a large amount of data based on qualitative and quantitative data. However, this can be very laborious in practice. Thus, the application of GenAI can support UX research activities. This paper provides a practical perspective on this topic. Based on previous studies, we present different use cases indicating the potential of GenAI in UX research. Moreover, we provide insights into an exploratory study using GenAI along an entire UX research process. Results show that Large Language Models (LLMs) are useful for various tasks. Thus, the research activities can be carried out more efficiently. However, the researcher should always review results to ensure quality. In summary, we want to express the potential of GenAI enhancing UX research.

## I. INTRODUCTION

A good UX fosters overall customer satisfaction and, in consequence, loyalty [1][2]. One of the main tasks of UX researchers is to evaluate the subjective impression of users towards products and to trigger improvements to ensure that they stay competitive over time. Users' subjective impressions can be broken down into three phases: (1) the anticipated use, (2) the actual use, and (3) after use. More precisely, it refers to the expectations before the use, the experience during, and the experience after using a product [3]. For this, different evaluation methods are applied in practical research.

A popular method to evaluate specific aspects of UX related to system or product quality is usability testing. Different testing procedures are available [4][5], but the general idea is similar. Test tasks are formulated, and participants are observed while working on them. Usually, participants are instructed to comment on what they do or want to achieve (thinking aloud). The researcher takes notes, or the comments are recorded. As a result of such a usability test, typically, an analysis of the comments of the participants together with some metrics (normally task completion rate and task completion time) is reported.

Usability tests provide detailed insights into problems and potential product improvements. But they require a lot of effort. For example, the preparation of good test scenarios and instructions for participants, the analysis of a large number of comments per participant, and the preparation of the insights for the test report. Thus, typically, only a small number of testers can be recruited. A survey with experts that conducted formative usability tests [6] found that 82% invited less than 15 testers and the overall median of test participants was 10 (similar results were reported in [7]).

Another quite popular method is the usage of surveys [8]. Compared to usability tests, it is very easy to recruit larger samples of participants in such studies. A survey can be sent by email to a large group of customers, posted on social media channels, or incorporated by a feedback button into the product's user interface.

UX and Usability surveys typically collect subjective impressions of users that are a mix of structured and unstructured data. Examples of structured data are demographic data (e.g., age, gender, profession), usage data (e.g., frequency of use, experience with the product), and scores from standardized UX and Usability questionnaires (e.g., SUS [9], UEQ [10], UMUX [11], or VISAWI [12]). Structured data are easy to analyze with standard statistical methods. Demographic or usage data often help to interpret the results of UX scores from questionnaires, for example, a higher frequency of use is often associated with higher UX ratings [13][14].

It is a common practice to allow participants to comment on the product by one or several text entry fields. Such comments create additional insights into why participants have answered a certain way and can often contain ideas for improvements. The collected comments are usually much shorter than transcripts from a usability test, and only a fraction of the participants fill in comments [15]. But their number can still be huge. If you get 10,000 responses and just 10% of them fill in some text, you still have 1,000 texts to analyze. This makes it clear that anything that helps to avoid a purely manual interpretation and analysis of comments is highly welcome.

In practice, it is common to conduct a mixed-method approach to gather comprehensive insights into the users' subjective impressions. Therefore, different methods, such as usability testing and surveys, are applied together. Thus, research activities require the realization of different tasks, such as the creation of textual artifacts for the preparation and execution of the activity. Examples include instructions or task descriptions for usability tests, personas resulting from user research, or the interpretation of users' comments. Moreover, both qualitative and quantitative evaluation results are collected. This textual data must be processed. LLMs can naturally support tasks, such as creating and interpreting textual data. Thus, it is interesting to know how they can be used in typical user research activities to save effort.

This paper is structured as follows: Section II explains the specific objective of the research, including the research questions. Afterward, the paper structure is aligned with the four research questions. Section III, IV, and V provide insights into applying GenAI in UX research referring to RQ 1 - 3. Section VI presents a practical approach applying different methods supported by GenAI. We summarize answers to the research questions in the discussion in Section VII. This article ends with a conclusion and outlook in Section VIII.

## II. RESEARCH OBJECTIVE AND RESEARCH QUESTIONS

The rapid development of GenAI techniques offers many possibilities to support UX research activities. Of course, it is not possible to oversee all the possibilities and risks associated with this new technology at the moment. However, some promising application areas can already be identified and are discussed in this paper. Against this background, we specifically address the following research questions:

- *RQ1: How can GenAI assist UX researchers in categorizing and summarizing user comments?*
  We investigate the potential of ChatGPT to analyze user comments from surveys and transcripts from usability tests using existing survey data from a larger study.

- *RQ2: How can GenAI be used to assist UX researchers in the preparation of research and design activities?*
  Several UX activities require the preparation of instructions for the participants (e.g., task descriptions for usability tests) or some artifacts used for design activities (e.g., personas). We investigate how GenAI can support UX researchers in those tasks.

- *RQ3: Can GenAI be used to develop a deeper understanding of the concept of UX?*
  We investigate how LLMs can be used to analyze the semantic textual similarity of items and scales from UX questionnaires and how these results can be used to plan UX evaluations.

- *RQ4: How can GenAI be integrated with existing methods in a typical UX evaluation scenario?*
  We describe the results of a UX evaluation based on a mixed-method approach that evaluates UX before (expectations concerning UX), during (analyzing Thinking-aloud comments from a product testing), and after (UEQ [10] results) use.

## III. GENERATIVE AI-BASED COMMENT ANALYSIS

There are several established methods to analyze and communicate user comments. A semantic grouping of comments into categories, together with the number of comments per category, can be communicated. Alternatively, a short textual summary of the main points mentioned by the users can be created. Other common methods are sentiment analysis [16] or word clouds [17][18]. LLMs can be used to assist UX researchers in all these tasks.

We will discuss the first two methods (sentiment analysis with LLMs is an already established method, and the creation of word clouds does not really need GenAI capabilities) using an example of a survey that collected feedback concerning PayPal. For details of the study, see [19]. The comments are answers to the questions "What do you particularly like about PayPal?" (positive comments) and "Which aspects do you find particularly bad about PayPal?" (negative comments). Overall there were 113 negative comments and 105 positive comments.

A ChatGPT (version 4.0) prompt consisting of general instructions followed by a numbered list of all positive (negative) comments from the survey was used. The instruction had the following content:

*The numbered list below contains user comments concerning PayPal. Ignore negative comments. Categorize the comments. One comment can belong to more than one category. Create a short name for each category that contains less than two words. Create a summary for each category in a very short sentence. List all categories. Show the short name of the category first. Then, in a new line, the summary of the category. Then the list of comments belonging to that category. Combine the short summaries from the 5 most frequent categories in a short text.*

Participants entered negative comments in some cases to the question "What do you particularly like about PayPal?" These are filtered out by the "Ignore negative comments" statement. To analyze the negative comments, this part was replaced by "Ignore positive comments."

The prompt splits the task of creating a classification and a summary into a simple procedure. Firstly, categories are determined and described. Since the categories and the assigned statements are listed in the result, a manual quality check of the categorization is easy. Secondly, the short statements are combined into a summary. Of course, the number of categories is not known beforehand. However, this number can be determined by simply running the prompt several times with different values for this number, which was, in fact, used to come to our decision to use 5 categories. The output indicates if there are not enough comments per category or if the categories are too abstract or do not clearly separate between comments.

In our example, the following five categories and short texts were produced for the positive comments:

- Ease of Use: Users appreciate the simplicity and intuitiveness of PayPal. (53 comments)
- Speed: Users value the quickness of transactions with PayPal. (21 comments)
- Security: Users feel secure using PayPal for their transactions. (13 comments)
- Integration: Users like how PayPal integrates with other platforms and services. (11 comments)
- International/Currency Features: Users appreciate the ability to handle international transactions and currency conversions. (7 comments)

For the negative comments, the result is as follows:

- Security Concerns: Users have concerns about the security of PayPal. (17 comments)
- Difficulties with Transfers and Withdrawals: Users find it difficult and time-consuming to transfer or withdraw money. (14 comments)
- High Fees: Users are unhappy with the high fees charged by PayPal. (9 comments)
- Poor Customer Service: Users find PayPal's customer service to be lacking and unhelpful. (8 comments)
- User Interface Issues: Users find PayPal's user interface confusing and difficult to navigate. (4 comments)

As a positive summary, the following text was created: *Users appreciate PayPal for its ease of use, quick transaction speed, security features, integration with other platforms, and its ability to handle international transactions and currency conversions* and the negative summary was *Users have expressed dissatisfaction with PayPal due to high fees, poor customer service, difficulties with transfers and withdrawals, issues with the user interface, and concerns about security. These areas represent the most frequent categories of negative feedback about the platform.*

Since ChatGPT is not deterministic, it must be checked how reliable such a generated categorization or summary is. If the same prompt is used twice, the results will usually differ to some extent. However, since the effort to run the same prompt several times is minimal, this can be checked easily. In our example, repetitive runs produced only small deviations, for example, "Users feel secure using PayPal for their transactions" versus "Users appreciate the security features of PayPal, including protection of personal details" or "Users find the ability to handle and convert different currencies useful" versus "Users find PayPal's ability to handle, convert, and transfer different currencies beneficial".

Another question is how well the result describes the content of the comments. Of course, there is no "optimal" solution to such tasks. Different human experts will also produce slightly deviating results. The validity of the result can be checked by manually analyzing a small subset of comments and comparing them to the classification of ChatGPT. To get a deeper understanding, we did (which is unrealistic for real use cases with thousands of comments) a full analysis of all comments by a human expert. The classification was identical in 88% of the cases.

Comments differ concerning content and length between different data collection mechanisms. If, for example, a product survey is sent to users via an email campaign, then users answer the questions in retrospect. In such cases, comments are often relatively short and concentrate on the product's general strengths and weaknesses. If a survey is opened over a feedback link inside the product, users answer the questions in their actual working context. Many comments will refer to the screen on which the feedback button is clicked and will be quite concrete (refer to a UI element or feature on the screen). Since the users assume that it is clear what they are referring to, the screen itself is typically not mentioned in the comment. Such comments are hard to interpret if the context is not known. Thus, in such cases, a human UX expert who knows the application well obviously has a massive advantage over an LLM. In each case, a more detailed analysis of the comments is required, for example, a pre-processing by analyzing comments collected for separate screens or adding some context information to the prompt.

In our example, a common product (PayPal) was investigated. Therefore, it was not necessary to provide additional context information; the product name was sufficient. In practical applications, the LLM will typically not have any information about the product (it will not be in the training set). It is recommended to add several sentences to the prompt that explain the product's main use cases.

IV. CREATION OF PERSONAS

Another area where LLMs can be quite helpful in UX research is the creation of texts used for research actions (we will show in Section VI how task descriptions for usability tests can be created) or for the communication of research results. In this section, we will show how the creation of personas can be supported by LLMs.

Personas [20][21] are fictional characters that represent typical target groups of users. They are created based on research results concerning the actual or potential users. Personas are merely a communication tool to help designers or developers empathize with users and consider their needs adequately. It is a matter of taste which information a persona description contains. There are several templates available that contain a pre-selection of categories and a way to visualize the persona description. Typical examples are demographics (name, age, gender, family status, etc.), goals and actions associated with the product (user journey, typical tasks, usage scenarios, etc.), skills (education, degrees, knowledge concerning product or work area, etc.), interests (for example, hobbies), personal goals or values (lifestyle, environment, etc.) and often a photo. Most of this information is not relevant to the product design decisions. They are added to make the persona "real" [20]. The underlying assumption is that it is easier to discuss if a product fulfills the needs of a real person (described by the persona) than an abstract "user", where every member of a design team may have different ideas about ([20]).

The goal of this paper is not to discuss the pros and cons of personas in design. They are a frequently used tool, and we show how easily LLMs can generate realistic persona descriptions. As described above, the core of a persona is research about the user group. We take as an example a developer of cloud applications that are used to enhance an existing ERP system in a company.

The following prompt contains the core information about such a developer role: *In UX research, personas are often used in user interface design to communicate and discuss design decisions. Create a persona for a software developer. The person works for a medium-sized company and is a member of a small team of software developers. The company uses SAP as its major ERP system. The development team creates cloud applications that enhance this ERP system with additional functionalities. Some key challenges in this job are the technical complexity of the applications and the need to get a deep knowledge of the structure and existing interfaces of the ERP system. Can you create a persona description? Please provide, in addition, a quote that could realistically come from the persona, which sums up what matters most to them in relation to the product or service.*

The resulting persona was described as follows:

- Persona Name: John Smith
- Background: John is a 35-year-old software developer who has been working in a medium-sized company for the last 7 years. He holds a Bachelor's degree in Computer Science and has a strong background in cloud computing and ERP systems.
- Role: John is a member of a small team of software developers. His main responsibility is to develop cloud applications that enhance the company's major ERP system, SAP, by adding additional functionalities.
- Key Challenges: John often struggles with the technical complexity of the applications he is developing. He also finds it challenging to gain a deep understanding of the structure and existing interfaces of the ERP system.
- Goals: John's primary goal is to develop robust, efficient, and user-friendly applications that can seamlessly integrate with the existing ERP system. He also aims to continuously improve his technical skills and understanding of the ERP system to overcome his challenges.
- Skills: John is proficient in several programming languages, including Java, Python, and C++. He also has a good understanding of cloud computing platforms, especially AWS and Google Cloud. He has a basic understanding of SAP but is still working on improving his knowledge in this area.
- Behaviors: John is a problem-solver who enjoys tackling complex technical challenges. He is a quick learner and is always eager to learn new technologies and tools that can help him in his job. He prefers to work in a collaborative environment and values clear communication within his team.
- Needs: John needs clear documentation and training materials to help him understand the ERP system better. He also needs a supportive work environment where he can freely discuss his challenges and get help from his team members.
- Quote: "Understanding the ERP system is just as important as writing good code. We need to ensure that our applications not only work well but also integrate seamlessly with the existing system. It's a challenge, but it's also what makes our job interesting and rewarding."

There are some remarkable points concerning this persona description. It contains several elements that are not explicitly mentioned in the prompt but inferred from this information. For example, "programming skills in Java, Python, and C++", or "understanding of AWS and Google Cloud". Also, the age (35), working experience (7 years), and degree (bachelor) fit together. The description under "Behaviors" fits well for a young motivated developer. Overall, the generated description matches the small amount of information contained in the prompt quite well and is a realistic (a bit stereotypical) description of a person.

It was already noted that personas often contain stereotypes [22]. This is true for personas created by UX researchers and even more true for ChatGPT-created personas. If we run, for example, the prompt described above 10 times, then all 10 personas are male and between 32 and 35 years old. If this is really a problem, it is a matter of taste and assumptions on how personas should be used in a design process. Stereotypical personas are, on the one hand, more realistic [20]; on the other hand, [23], they can cause wrong conclusions and predictions in the design team. However, at least concerning demographics contained in a ChatGPT-created persona, this can easily be avoided by explicitly defining this information in the prompt.

V. ANALYZING SEMANTIC SIMILARITY OF ITEMS AND SCALES

This section shows how LLMs can be used to get deeper insights into semantic similarities of UX items. We focus on the application aspects. More details concerning the methods can be found in [24][25].

The most common way of measuring UX is the usage of standardized questionnaires in surveys [26][27]. UX has many facets, thus a single questionnaire can not cover the whole concept [28]. Many different UX questionnaires are available, and each one focuses on different UX aspects. Thus, it depends on the evaluation scenario which questionnaire is suitable [29][30].

A UX questionnaire consists of different items and scales [29][30]. However, as the example items in Table I show,

nearly identical items can be assigned to differently named scales. Conversely, scales with highly similar or even identical names can measure semantically different concepts. For example, AttrakDiff [31] and UEQ [10] both contain a scale named *Stimulation*. However, *Stimulation* in the sense of the UEQ refers to an interesting and stimulating experience. *Stimulation* in the sense of the AttrakDiff2 contains, in addition, the aspect that the design of the product is creative and innovative [8].

TABLE I. EXAMPLE OF SIMILAR ITEMS ASSIGNED TO DIFFERENTLY NAMED SCALES [30].

| Item | Scale | Questionnaire | Source |
|---|---|---|---|
| The system is easy to use | Likeability | SASSI | [32] |
| I thought the system was easy to use | Usability | SUS | [9] |
| This system is easy to use | Overall | UMUX | [11] |
| It was simple to use this system | System Usefulness | PSSQU | [33] |

Previous research concerning the dependency of UX scales mainly focused on an analysis of correlations between scales [29][30][34]. Other studies analyze the semantic textual similarity by applying NLP techniques [35][36], i.e., Sentence Transformers, to analyze the similarity of the encoded textual items in a vector space.

LLMs are good candidates for performing an analysis of semantic similarity between items and scales of UX questionnaires. They use word embeddings to represent the semantics of texts. Against this background, we applied ChatGPT-4 to (1) (re-) construct common UX concepts based on a set of UX items (see V-A), (2) detect and match suitable measurement items based on semantic textual similarity (see V-B), and (3) uncover the semantic textual similarity among the measurement items (see V-C).

*A. Detecting a Semantic Structure on Items*

We created an item set containing 408 items from 19 UX questionnaires (for construction details, see [24][25]). A series of six ChatGPT prompts are formulated and sequentially applied that ask ChatGPT to classify the items into categories and more fine granular sub-categories (see [24][25] for the detailed formulation of the prompts). As a result, the LLM generated six main topics and 16 subtopics:

- **System Usability and Performance**: Ease of Use — Efficiency and Speed — Functionality and Flexibility
- **User Engagement and Experience**: Engagement Level — Aesthetics and Design — Confusion and Difficulty
- **Information and Content**: Clarity and Understandability — Relevance and Utility — Consistency and Integration
- **Website-specific Feedback**: Navigation and Usability — Trust and Security — Aesthetics and Design
- **Learning and Adaptability**: Learning Curve — Adaptability
- **Overall Satisfaction and Recommendation**: Satisfaction — Recommendation

The full classification can be found in [24][25]. As an example, we show the top five items of the sub-category *Efficiency and Speed*:

1) The interaction with the system is fast.
2) The system responds too slowly.
3) This software responds too slowly to inputs.
4) The speed of this software is fast enough.
5) Has fast navigation to pages.

The generated topics refer to both pragmatic and hedonic properties, but they are rather broad and sometimes look strange from the perspective of a human UX expert. For example, the subtopic *Navigation and Usability* referring to a pragmatic value and *Aesthetics and Design* referring to a hedonic value are summarized under the main topic *Website-specific Feedback*. Most of the generated sub-topics can be easily related to established UX concepts, for example, a larger set of UX aspects used in several studies [30].

In another prompt, we inserted the UX quality aspects from [30] and asked ChatGPT to compare them with its own generated sub-topics. The comparison is shown in Table II:

TABLE II. COMPARISON OF EXISTING UX QUALITY ASPECTS BY [30] AND AI-GENERATED TOPICS.

| (#) | UX Quality Aspects | AI-generated Sub-Topics |
|---|---|---|
| (1) | Perspicuity | Ease of Use — Learning Curve |
| (2) | Efficiency | Efficiency and Speed |
| (3) | Dependability | Consistency and Integration |
| (4) | Usefulness | Functionality and Flexibility—Relevance and Utility |
| (5) | Intuitive use | Ease of Use |
| (6) | Adaptability | Adaptability |
| (7) | Novelty | - |
| (8) | Stimulation | Engagement Level |
| (9) | Clarity | Clarity and Understandability |
| (10) | Quality of Content | Relevance and Utility |
| (11) | Immersion | Engagement Level |
| (12) | Aesthetics | Aesthetics and Design |
| (13) | Identity | - |
| (14) | Loyalty | Loyalty |
| (15) | Trust | Trust and Security |
| (16) | Value | Perceived value |

Results show that most AI-generated sub-topics are named differently but can be allocated to established UX quality aspects. *Novelty* and *Identity* were not classified, but this was caused by the fact that most of the questionnaires in our list did not contain corresponding items.

To sum up, ChatGPT-4 generates a comprehensive overview of topics and subtopics based on UX measurement items. Moreover, both pragmatic and hedonic topics are contained. The different items almost completely match the respective topic. Thus, LLMs are useful in generating topics and thus can be used to investigate the semantic structure of sets of items.

## B. Finding Suitable Items for Ad Hoc Surveys

Standardized UX questionnaires offer many advantages compared to ad hoc defined surveys [8]. However, it is also not uncommon to use surveys in an exploratory way to get insights into the opinions of users. Such ad hoc surveys often contain open-ended questions and some closed questions to get some rough scores concerning single UX aspects. Often single questions from standardized questionnaires are reused for this purpose. In this section, we show how LLMs can be used to determine suitable items.

ChatGPT-4 was applied to the data set of 408 items described above. The following exemplary prompt was used to find items that describe the aspect of *Usefulness* (can be adapted to other UX qualities by changing the bold part):

*Below there is a list of statements and questions related to the UX of a software system.* **Select all statements or questions from this list that describe whether the software system is useful or not.** *List these statements or questions. Start with those statements and questions that describe this best.*

As a result, ChatGPT identified 15 items. The top 5 are shown below.

1) The software helps me to complete my work tasks better than expected without extra effort.
2) With the software, I can sometimes even exceed my desired goals without any extra effort.
3) The software allows me to increase the quality of my work without any extra work.
4) The software offers me all the possibilities I need to work on my tasks.
5) The software is tailored to the tasks I need to work on.

The selected items fit pretty well with the intention formulated in the prompt. Thus, if an adequate list of items is available, it is easy to filter out candidates who match a given intention. For details and adjustments of the described prompt to other UX aspects, see [25]. This paper shows that, with a few exceptions, the selected items aligned well with the UX quality specified in the prompt. However, in this use case the intention is to detect one or a few items that describe a certain aspect of UX, thus researchers will pick items from the top of the list. Thus, ChatGPT can be used to select items from a candidate list that fit a certain research objective.

## C. Investigating the Semantic Similarity between UX Concepts

In this section, we want to provide an insight into how effective GenAI is in measuring semantic similarity. Previous research has shown that innovative NLP techniques can be applied to compare the semantic similarity based on the encoded textual items in a vector space [35][36]. As GenAI is a sub-field of NLP, the semantic textual similarity analysis was conducted by applying ChatGPT-4.

A second item pool was used for this study. A list of 40 UX questionnaires [28] was analyzed, and artificial items with a highly standardized format were created. All positive adjectives from both existing semantic differential scales and statements in UX questionnaires were extracted. Based on the positive adjective, items were generated applying the same structure: *"I perceive the product as ¡adjective¿"*. This results in a data set of 135 artificially generated items. We refer to [8][37] for a similar technique.

We use ChatGPT to investigate how the 135 artificial items relate to existing UX concepts. Therefore, we applied a generic prompt with an instruction and an explanation regarding common UX concepts/quality aspects. In the following, an exemplary prompt concerning *Usefulness* is illustrated:

*Below, there is a list of statements related to the user experience of a product.* **Select all statements from this list that describe that users perceive the product as useful.** *List these statements or questions. Start with those statements and questions that describe this best.*

After the prompt, we inserted the artificially generated item pool. The part of the prompt marked in **bold** was adapted to the respective UX quality aspect. The remaining part of the prompt stayed stable. In summary, prompts for the UX quality aspects *Learnability*, *Efficiency*, *Dependability*, *Stimulation*, *Novelty*, *Aesthetics*, *Adaptability*, *Usefulness*, *Value*, *Trust*, and *Clarity* were applied. The respective prompt for each quality aspect runs three times. Only the items assigned to the respective UX quality aspect in all three runs were included. This results in an overview with semantically similar adjectives representing the respective UX quality aspects. For detailed results, we refer to [25].

Such an analysis can provide insights into the semantic overlap of common UX concepts. We see, for example, that *Novelty* and *Stimulation* share several assigned adjectives, thus there is some semantic overlap. This corresponds well to the realization of these concepts in UX questionnaires. In the AttrakDiff [31], these aspects are combined in one scale, while they form two separate scales in the UEQ [10]. There is also the expected close connection between the pragmatic qualities *Learnability*, *Efficiency* and *Dependability*. We can also see that *Clarity* and *Value* are somewhere in between the pragmatic qualities and *Aesthetics*, which is also known from empirical research [38][39]. However, many more dependencies become visible in the graphic. Thus, such an analysis can provide insights into otherwise hidden semantic overlaps of common UX concepts.

An analysis, as described above, can help UX researchers develop a deeper understanding of common UX scales and concepts and their semantic overlap. It can also help interpret the results of UX surveys that use corresponding UX scales for measurement.

## VI. Exploratory Study for the Investigation of a Generative AI-supported UX Evaluation Scenario

The previous sections showed the potential of GenAI for different tasks in UX research. This section will provide a practical example of how GenAI can be applied in a research process. We conducted an exploratory study based on a mixed-method approach, gathering qualitative and quantitative data. The study aimed to determine the extent to which GenAI is a useful enhancement in the practical UX research process. In the following, the study approach is described.

*A. Study Approach*

The study is designed to present an entire UX research process. All relevant aspects of UX are therefore intended to be examined. This results in three perspectives:

1) **Anticipated Use**: UX impressions and expectations towards a product before the use [40]
2) **Actual Use**: The momentary experience during the interaction with the product.
3) **After Use**: The retrospective experience after the usage of the product.

Thus, the methodological approach is broken down into these three parts. It is important to note that all three parts of this study are based on the theoretical foundation of the UEQ concerning the construct of UX [10]. Thus, the common ground is the six UX scales and their respective items. For details, we refer to [10].

We applied the UEEE method developed by [40] regarding part *(1)* **Anticipated Use**. Moreover, we conducted a usability test concerning part *(2)* **Experience during Use**. In particular, the Thinking-Aloud method based on AI-generated user tasks was performed. This results in a textual data pool of 115 pages and approximately 43.750 words. Afterward, the resulting user protocols were analyzed using ChatGPT-4. In relation to part *(3)* **Actual Use**, we applied the User Experience Questionnaire (UEQ) by [10]. The approach is visualized in Figure 1.

The social media platform **Instagram** was analyzed as an evaluation object. We conducted a convenience sample with a total of 30 study participants. The study was conducted in German. In the following, the three performed parts are further described.

*1) User Expectations:* User Experience Expectation and Evaluation (UEEE) [40][41] is a method to evaluate user expectations uniformly and efficiently. The method is based on the same idea as card sorting methods, such as *Product Reaction Cards* [42]. The researchers define a set of adjectives describing the respective product. These adjectives represent the expectations towards the product. The adjectives are displayed to the study participants in a tool described in [40]. The study participants are asked to evaluate these adjectives based on their expectations concerning the product. Therefore, the adjective can be classified into four categories (*unimportant, rather unimportant, rather important, important*). For this study, we used the positive adjectives of the 26 items of the UEQ.

Regarding the evaluation, study participants received a brief introduction and explanation of the tool and the evaluation object. Afterward, they had to classify the adjectives based on their expectations of using Instagram.

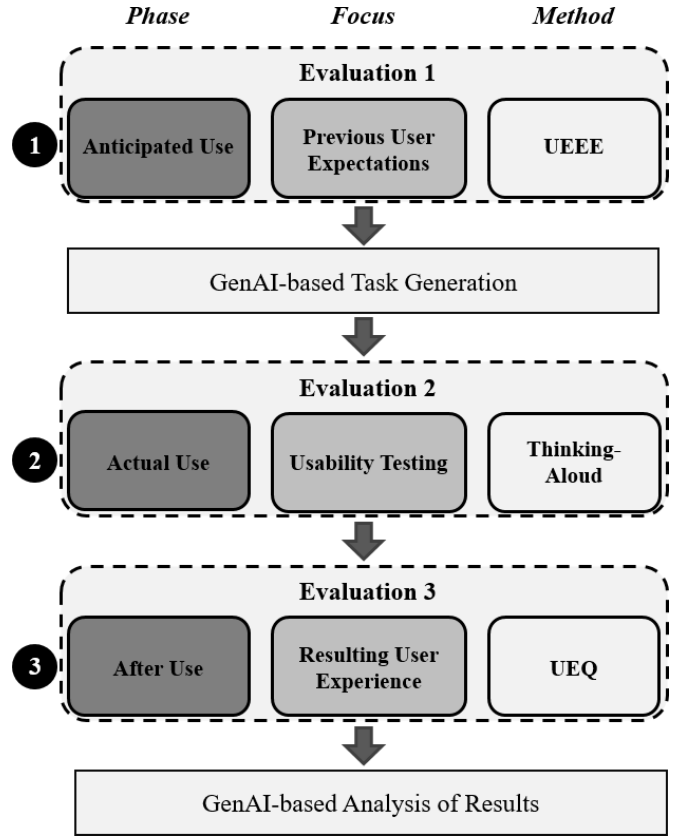

Figure 1. Methodological Approach of the Exploratory Study.

*2) Generative AI-based Task Generation:* We used ChatGPT-4 to generate tasks concerning usability testing. We applied four prompts to specify the tasks for usability testing. Each task refers to one UX scale regarding the UEQ [10]. Thus, this ensures that the user protocols regarding the Thinking-Aloud relate to the relevant UX factors. The four prompts and the final tasks are provided in Appendix A and B.

ChatGPT's ability to generate tasks results from the fact that Instagram is a well-known product. In the training phase of the LLM, there were obviously many texts concerning the usage of Instagram. In practice, usability tests are typically performed for new products, and such information is not available for the LLM. However, existing product documentation, specifications, or other sources of information can be added to the prompt in such cases to provide the required information.

*3) Actual Use:* To measure the experience during use, we conducted usability testing. More precisely, we deployed Thinking-Aloud. For this, the study participants were introduced to the UX test scenario. Moreover, they received the six AI-generated tasks to complete using Instagram.

*4) User Experience:* In the third part, we evaluated the actual UX. Therefore, various methods for measuring UX can be found in the scientific literature. The most established way is the use of standardized questionnaires.[26]. We implemented

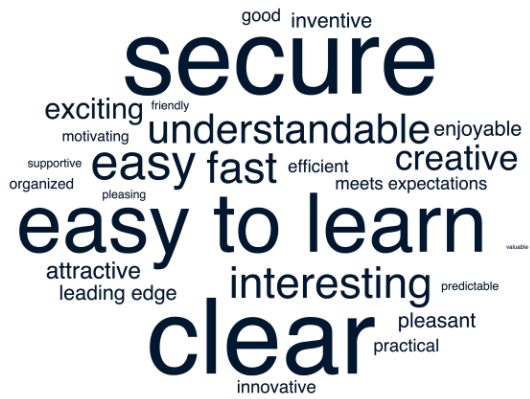

Figure 2. Word Cloud Visualizing the Expectations of Participants.

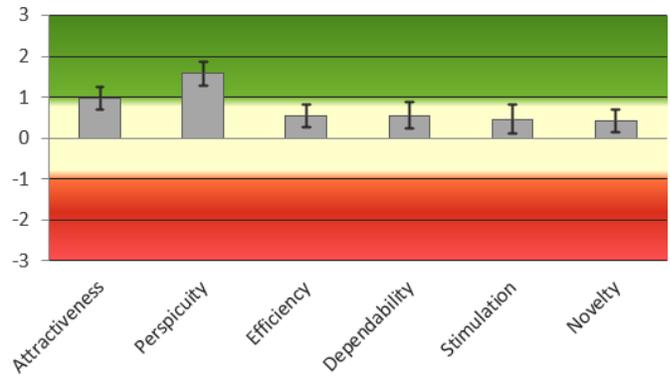

Figure 3. Scale Scores and Confidence Intervals for the UEQ.

the UEQ by [10] to examine the UX.

The results are shown below. Due to paper restrictions, we focus on the relevant results regarding GenAI.

### B. Results

*1) Anticipated Use – UEEE Results:* Figure 2 shows a word cloud that visualizes how often an adjective was assigned to the category *important*. The bigger an item is displayed, the more participants consider it important for the product. The word cloud shows that users have high expectations concerning security, perspicuity (easy to learn, clear, understandable), and stimulation (interesting, exciting).

*2) Actual Use:* As shown in Section V-A based on [24][25], GenAI is useful in (re-)constructing UX factors. Thus, we further applied ChatGPT-4 to identify UX factors by analyzing the user protocols conducted with the Thinking-Aloud method. In particular, we consolidated the qualitative textual data based on the different tasks. This resulted in six consolidated protocols. We used ChatGPT-4 to analyze data for each task to identify key topics concerning the UX. The identified topics and descriptions are shown in the following:

- **Ease of Use and Intuitiveness:** Many users found Instagram's story creation features to be intuitive and easy to navigate. Features like adding location tags, filters, hashtags, and polls were generally considered user-friendly.
- **Feature Discovery:** Some users indicated challenges in discovering specific features or functions within the app, suggesting that while the app is generally easy to use, certain functionalities may not be immediately obvious to all users.
- **Personalization and Creativity:** The availability of various filters, stickers, and the ability to add music was positively received, offering users creative ways to personalize their stories.
- **Information Clarity:** There were mixed responses about the clarity of information presented in the app. Some users found it straightforward to understand and use different features, while others struggled with specific aspects like changing colors or understanding the purpose of certain icons.
- **Learning Curve and User Guidance:** New users or those less familiar with the app's functionalities experienced a learning curve. However, once familiarized, they found the app easier to navigate.
- **Functionality and Performance:** Overall, the app's functionalities were well-received. However, some users suggested improvements, like better filter variety or more options for customizing location stickers.
- **Accessibility and User Experience Variability:** Different users had varying experiences based on their familiarity with Instagram, indicating that user experience can vary significantly depending on the user's prior exposure to the app.

*3) After Use – UEQ Results:* Figure 3 shows the results of the UEQ. As we can see, there is a high rating for the scale *Perspicuity* and also a positive rating for the scale *Attractiveness*. The other scales show only moderate positive ratings. Of course, the confidence intervals are relatively big, based on the fact that we had only 30 participants. Thus, we should interpret the result with care. But if we compare this to the expectations evaluated before, we see that the expectations concerning *Perspicuity* are fulfilled, while this is not the case for *Stimulation*.

*4) Matching Qualitative and Quantitative Evaluation Results:* In the last step, we applied ChatGPT-4 to match qualitative and quantitative evaluation results. In particular, we aimed to match the textual data of the user protocols with the quantitative results of the UEQ and the UEEE. Thus, we provide a combination of quantitative scale values and qualitative text content. Two perspectives were provided for this.

1) **Quantitative Data Perspective**
2) **Qualitative Data Perspective**

Concerning (1) **Quantitative Data Perspective**, we aimed to identify the number of relevant statements in the textual data (user protocols) related to the respective items. In relation to (2) **Qualitative Data Perspective**, we aimed to classify relevant statements to the different items. Thus, each scale and its items consist of relevant textual data. We analyzed the textual data regarding the UX factors *Attractiveness* and *Stimulation*. We applied five prompts to ChatGPT-4. Each prompt was adjusted for the respective UX factor. In the following, we present the respective prompts and results. Due to paper restrictions, we conducted results referring to the UX factor *Attractiveness* and *Stimulation*.

The first prompt is as follows: ***Prompt1:*** *We performed usability testing regarding Instagram. For this, we generated a task. The task refers to the UX factor Attractiveness and its positive items.* **Inserted definition and items**. *The attached user protocol file contains 30 answers given to the task. As the tasks relate to the specific factor of Attractiveness and its items, it should be possible to match the textual data with the items. I want you to analyze all statements. Match all relevant statements to the respective items.*

We aimed to provide an initial classification of the statements for the items of Attractiveness. As a result, ChatGPT-4 generates a first classification. For this, the LLM generalized the analyzed textual data. Results are shown in the following. We only present an example of the first two item pairs:

- *Attractiveness:* **Annoying / Enjoyable**
  Annoying: Frequent complaints about too much advertising, unorganized feeds, and overwhelming amounts of content were common, making the experience less enjoyable for some participants (e.g., Participants 1, 7, 14, 22).
  Enjoyable: Others found the diversity in content and the presence of videos and stories to be enjoyable, citing the variety as an "Eyecatcher" and appreciating the personalized feed (e.g., Participants 2, 5, 21).

- *Stimulation:* **Not Interesting/Interesting**
  Interesting: Participants who encountered content aligned with their interests found the Reels feature interesting. For example, Participant 6 appreciated the ease of scrolling and discovering content, though they wished for better control over video pausing.
  Not Interesting: Feedback from Participant 2 and Participant 9 suggested that not all content was appealing, with some users finding the Reels not tailored enough to their preferences or too similar to content from other platforms like TikTok.

Moreover, we want to provide both quantitative and qualitative perspectives as described before. Therefore, we applied **prompt2 and 3**.

***Prompt2:*** *We want to show a quantitative perspective. Therefore, we want to illustrate the number of statements referring to each of the respective items. Can you analyze and classify all relevant statements to the six items?*

***Prompt3:*** *Furthermore, we want to show a qualitative perspective. Therefore, we want to illustrate the relevant statements referring to each of the respective items. Can you analyze and extract all relevant statements from the file and sort them into the six items?*

As a result, ChatGPT analyzed the quantity *(n=x)* of suitable statements concerning the items. Moreover, original statements were extracted. Due to paper restrictions, we only illustrated the first original statement extracted by ChatGPT-4 for each item.

*Attractiveness:*

- **Annoying / Enjoyable**: *n=7*
  "I think they are such eye-catchers." (Participant 2)
- **Bad / Good**: *n=4*
  "The design is clear and I can quickly see what I like." (Participant 8)
- **Unlikable / Pleasing**: *n=3*
  "Everything is very straightforward, clear and simple." (Participant 3)
- **Unpleasant / Pleasant**: *n=3*
  "I think that's even better, because the dark design is set for me." (Participant 3)
- **Unattractive / Attractive**: *n=2*
  "Visually appealing in any case...I like round icons very much." (Participant 22)
- **Unfriendly / Friendly**: *n=2*
  "It is easy to scroll. It is understandable." (Participant 5)

*Stimulation:*

- **Inferior / Valuable**: *n=22*
  "I enjoy scrolling through the reels." (Participant 10)
- **Boring / Exciting**: *n=15*
  "But I find the reels very exciting. It motivates me to keep watching." (Participant 4)
- **Not Interesting / Interesting**: *n=25*
  "There are lots of interesting things coming up and that takes up a lot of my time in the evenings." (Participant 3)
- **Demotivating / Motivating:**: *n=18*
  "Sometimes you have to watch the reel twice. Maybe 3 or 4 until you finally stop at the right place." (Participant 6)

In the last step, we inserted **prompt4 and 5** to match the quantitative and qualitative evaluation results. The prompts are as follows:

***Prompt4:*** *In summary, we conducted a mixed-method approach. We examined the user expectations using UEEE, usability testing by applying Thinking Aloud, and the UX by applying the User Experience Questionnaire. Thus, we provide quantitative and qualitative results. In the following, quantitative results regarding the UX factor Attractiveness are shown.* **Inserted quantitative evaluation results** *The*

*first number is the quantity of how often the adjective was evaluated as important as expectation. The second number is the mean value from the questionnaire using a 7-point Likert scale. Compare the selected statements of the qualitative results with the quantitative evaluation results.*

**Prompt5:** *Bring the relevant statements analyzed from the user protocols classified to the respective items together with the quantitative results provided.*

After **prompt4**, the results were generally summarized by ChatGPT, similar to **prompt1**. Therefore, **prompt5** has been introduced to provide a more specific representation referring to the respective user statements. As a result, ChatGPT-4 considered both evaluation results and generated a matching summary for each item. Again, we only show the results of the first two items.

- *Attractiveness:* **Annoying / Enjoyable**
  Despite no initial expectations, users found Instagram enjoyable, particularly highlighting dynamic content, such as videos. This suggests Instagram's ability to deliver an enjoyable experience beyond users' anticipations.
- *Stimulation:* **Not Interesting/Interesting**
  The statements underscore the importance of content relevance and variety in making the platform interesting, which is mirrored in the high mention count and positive mean value, indicating a strong correlation between user interest and the Stimulation factor.

We conducted this procedure with the textual data from all six tasks. Finally, we combined all results from the exploratory study. Therefore, evaluation results were brought together in Figure 4 regarding the expectation (*Anticipated use*), experience (*After use*), and the number of statements (*Actual use*) identified before. The items are scaled from -1 to 1 concerning expectation and experience (see [40]). Moreover, the dot size linearly represents the frequency of comments. The different colors summarize the items of the respective UX factor.

The results show that items, such as *easy* or *understandable*, have a high rating of both expectations and experience, but few comments. This indicates that it is taken for granted and that this is confirmed afterward. In contrast, the number of statements is higher for the items rated with high expectations but a lower experience. This is plausible as usability test results are usually problem-centered and, thus, rather focus on negative aspects more.

To underline this, we performed the correlation between the three dimensions (see Table III). Expectation and experience strongly correlate. This means that Instagram generally seems to fulfill expectations well. The correlation between experience and the quantity of statements is strongly negative. Thus, the number of statements is greater when expectations are not met. In contrast, less is said when expectations are met. This emphasizes the fact that usability testing focuses on problems.

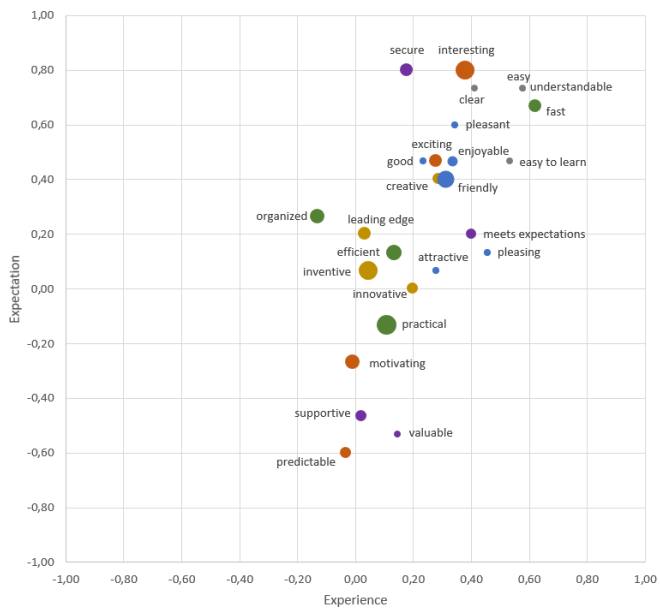

Figure 4. Comparison of Anticipated Use and Actual Use with Extended GenAI Results.

TABLE III. CORRELATION BETWEEN EXPECTATIONS, EXPERIENCE, AND STATEMENTS QUANTITY.

|  | Expectation | Experience | Statements Quantity |
|---|---|---|---|
| Expectation | 1 |  |  |
| Experience | 0.668819 | 1 |  |
| Statements Quantity | -0.29093 | -0.55583 | 1 |

To sum up, the LLM was useful in processing large text data. In particular, ChatGPT-4 was able to (1) generate a generalized summary, (2) identify, count, and extract suitable statements, and (3) match quantitative with qualitative evaluation results.

*C. Study Limitations*

Concerning this approach, we want to draw several limitations from a practical perspective. As shown, various tasks can be enhanced by applying LLMs. However, it depends on the level of information provided for the LLMs. This concerns both information supplied by prompts and accessed by the LLM itself. Moreover, the non-deterministic nature of LLMs must be mentioned. Thus, results can differ by applying the same prompt twice. Moreover, we did not follow any strategy in prompt engineering. Furthermore, the small sample size must be mentioned. Thirty study participants are just sufficient for the quantitative evaluation part.

VII. DISCUSSION

To sum up, we want to discuss the research questions based on the results. Therefore, we want to show the potential of applying LLMs for different tasks in the UX research process. Using LLMs makes it easy to adopt different perspectives in the research process. Thus, a large amount of data can be processed and analyzed quickly. This would not be possible with a manual analysis. Moreover, no specific strategy concerning prompt engineering was applied.

- *RQ1: How can GenAI assist UX researchers in categorizing and summarizing user comments?*
  Concerning the first research question, we focus on processing short text data. Therefore, user comments from a study were analyzed and summarized. Results show that the LLM was useful in analyzing and summarizing text data. However, it must be mentioned that the text data from surveys usually differ concerning content and length. The data is extremely heterogeneous. Thus, it is necessary to review AI-generated results by an expert.
- *RQ2: How can GenAI be used to assist UX researchers in the preparation of research and design activities?*
  Regarding the second research question, we provide insights into preparing a UX evaluation. In particular, we generated specific user tasks for the usability testing method Thinking-Aloud as well as personas. For this, the AI-generated user tasks were useful and specifically related to the research objective of the study. Moreover, ChatGPT-4 generated realistic personas. Remarkably, different aspects that were not described were generated. However, typical stereotypes were included.
- *RQ3: Can GenAI be used to develop a deeper understanding of the concept of UX?*
  Moreover, we provide insights into identifying a common ground concerning the concept of UX [24][25]. Therefore, we applied ChatGPT-4 to analyze the semantic textual similarity of UX measurement items from established UX questionnaires. Results show that GenAI was useful in (1) (re-) constructing common UX factors, (2) detecting suitable items, and (3) covering semantic similarity. Furthermore, adjectives were assigned to semantic similar UX concepts.
- *RQ4: How can GenAI be integrated with existing methods in a typical UX evaluation scenario?*
  Concerning the last research question, we conducted an exploratory study investigating users' subjective impressions comprehensively. We applied a mixed-method approach gathering both quantitative and qualitative data. Moreover, we used ChatGPT-4 for different tasks during the research process. In particular, we generated tasks for usability testing as well as analyzing results. Regarding the latter, we showed that quantitative and qualitative evaluation results could be matched. Moreover, improvement suggestions could be derived.

In summary, we state that using LLMs such as ChatGPT-4 is useful for various activities along the UX research process. Thus, the research process can be made more efficient. Above all, time is saved, and practical research processes can be enhanced. However, the researcher should always review results to ensure quality. As shown in this study, the following tasks concerning UX evaluation can be enhanced:

1) Analyzing both small and large text data
2) Creating personas
3) Analyzing semantic similarity
4) Generating user tasks
5) Matching quantitative and qualitative evaluation results

In the following section, we provide a conclusion and outlook.

## VIII. CONCLUSION AND OUTLOOK

This article provides insight into the potential of the new technology, GenAI, for UX research. We present different research activities and use cases supported and conducted by using the LLM ChatGPT-4. Results offer concrete usage scenarios for practice.

GenAI is a new and fast-developing area. New potential use cases evolve in short intervals, and it is currently not possible to judge their potential and Challenges definitively. It is important to notice that developed prompts supporting GenAI use cases must be constantly monitored. Moreover, a human expert should critically evaluate their results. Firstly, the underlying GenAI, or especially LLM, also develops over time. Thus, a new version may produce a different output to a prompt than expected. Secondly, since LLMs are non-deterministic, it is, in each case, a good idea to check how stable the output of a prompt is during several runs.

In conclusion, we want to clearly emphasize the potential to enhance UX research and its activities by applying GenAI. The UX research process and its tasks can be improved. Especially in practice, labor-intensive activities along the research process can be accelerated and supported. Moreover, a benefit can be shown in the derivation of results. In practice, different evaluation methods are often combined, resulting in various data. By using GenAI, the results can be matched with each other. This allows the researcher to differentiate from pure scale values. Thus, specific aspects for improving the evaluation object can be derived due to data matching. Conclusions providing both quantitative and qualitative results can be drawn.

This paper is a first step in providing a selected overview of how to apply LLMs in UX research. Future research should further consider the potential of enhancing UX research activities. Therefore, the different UX evaluation methods and their tasks should be investigated along the UX research process.

APPENDIX

*A:* Prompts for User Tasks in relation to the Thinking-Aloud

*prompt1:* Imagine that you want to analyze the user experience of Instagram. To do this, you will prepare a 12-minute user experience test with participants who already have an Instagram account. It is essential that the 6 UX factors and the items associated with them are taken into account and that they are explicitly addressed. For each factor, provide a singular and very detailed UX test task along with very specific instructions for the tester to accomplish that task. Write which goal is to be achieved with it. Dedicate around two minutes per task. In the end, there should be six very specific tasks with a clear aim. Testers should experience and evaluate all factors. The overall UX must be evaluated, not the content! In the end, the testers will be answering the UEQ. Here are the six factors contributing to the user experience (UX) and their evaluation scores: –inserted the six UX scales and their respective items.

*prompt2:* Make the tasks more specific, e.g., give a certain topic to search for.

*prompt3:* The user test will be conducted using the think-aloud method. Rewrite the instructions for the testers so that they report on their experiences as much as possible during the test. The test should still last 12 minutes and analyze the UX of Instagram, not the content of the app.

*prompt4:* The test should have a common thread (e.g., traveling, fitness, healthy eating).

*B:* User Tasks in relation to the Thinking-Aloud

1. **Attractiveness (2 minutes)**

- Task: Discover and interact with new travel content.

- Instructions: Go to the 'Explore' tab and type "travel" in the search bar. Choose and interact with three different posts (like, comment, or share) that you find most appealing. Look for posts with diverse content like landscapes, cityscapes, and cultural highlights.

- Goal: To evaluate how attractive and engaging the content discovery process is on Instagram.

2. **Efficiency (2 minutes)**

- Task: Post a travel-themed photo efficiently.

- Instructions: Select a personal travel photo from your gallery. Apply the 'Clarendon' filter, write a caption describing the location and your experience, add hashtags, such as TravelMemories and AdventureAwaits, and post it. Time how long it takes to complete this process.

- Goal: To assess the speed and practicality of posting content on Instagram.

3. **Perspicuity (2 minutes)**

- Task: Create and publish a multi-feature Instagram Story.

- Instructions: Create a Story featuring a travel photo. Add a geotag of the location, use the 'Poll' sticker to ask followers about their travel preferences, and add a relevant GIF. Ensure the Story is coherent and visually appealing.

- Goal: To evaluate how easily new users can understand and use various Story features.

4. **Dependability (2 minutes)**

- Task: Review and adjust account security settings.

- Instructions: Navigate to your profile, tap on the three-line menu, select 'Settings', then 'Privacy'. Set the account to private, then go to 'Security', enable two-factor authentication, and review active sessions. Revert any changes if desired after completing the task.

- Goal: To determine how dependable and user-friendly Instagram's privacy and security settings are.

5. **Stimulation (2 minutes)**

- Task: Engage with travel community content.

- Instructions: Follow the official Instagram account of a popular travel destination (e.g., @VisitParis or @ExploreNewZealand). Comment on their latest post with a question or a meaningful remark about the destination.

- Goal: To assess whether interaction with thematic content is stimulating and enjoyable.

6. **Novelty (2 minutes)**

- Task: Explore innovative content formats.

- Instructions: Go to the 'Reels' tab and watch three travel-related Reels. Look for content that showcases unique destinations, creative editing, or unusual travel experiences. Observe the layout and presentation of these Reels.

- Goal: To evaluate the creative and innovative aspects of Instagram's newer features.